\documentclass[final,5p,times,twocolumn]{elsarticle}

\usepackage[T1]{fontenc}
\usepackage[utf8]{inputenc}

\usepackage{amsmath}
\usepackage{amsfonts}
\usepackage{amssymb}
\usepackage{amsxtra}
\usepackage{array}
\usepackage{color}
\usepackage{dcolumn}
\usepackage{graphicx}
\usepackage{hepunits}
\usepackage{xspace}
\usepackage{CJKutf8}

\definecolor{purple}{rgb}{0.5,0,0.5}
\definecolor{blue}{rgb}{0.0,0,0.9}
\definecolor{prdblue}{rgb}{0.133,0.118,0.498}
\usepackage[colorlinks=true, pdfstartview=FitV, linkcolor=prdblue, citecolor= prdblue, urlcolor=prdblue]{hyperref}

\usepackage[mathscr,scaled=1.15]{urwchancal}
\DeclareFontFamily{OT1}{pzc}{}
\DeclareFontShape{OT1}{pzc}{m}{it}%
{<-> s * [1.15] pzcmi7t}{}
\DeclareMathAlphabet{\mathpzc}{OT1}{pzc}{m}{it}

\biboptions{sort&compress}

\journal{Physics Letters B}

\makeatletter

\setbox0\hbox{$\xdef\scriptratio{\strip@pt\dimexpr
    \numexpr(\sf@size*65536)/\f@size sp}$}

\newcommand{\scriptveryshortarrow}[1][3pt]{{%
    \hbox{\rule[\scriptratio\dimexpr\fontdimen22\textfont2-.2pt\relax]
               {\scriptratio\dimexpr#1\relax}{\scriptratio\dimexpr.4pt\relax}}%
   \mkern-4mu\hbox{\let\f@size\sf@size\usefont{U}{lasy}{m}{n}\symbol{41}}}}

\makeatother

\begin{document}
\begin{CJK}{UTF8}{song}

\begin{frontmatter}

\title{$\,$\\[-7ex]\hspace*{\fill}{\normalsize{\sf\emph{Preprint no}. NJU-INP 087/24}}\\[1ex]
Onset of scaling violation in pion and kaon elastic electromagnetic form factors}

\author[ECT]{Zhao-Qian Yao
       $\,^{\href{https://orcid.org/0000-0002-9621-6994}{\textcolor[rgb]{0.00,1.00,0.00}{\sf ID}}}$}

\author[ECT]{Daniele Binosi%
    $\,^{\href{https://orcid.org/0000-0003-1742-4689}{\textcolor[rgb]{0.00,1.00,0.00}{\sf ID}}}$}

\author[NJU,INP]{Craig D. Roberts%
       $^{\href{https://orcid.org/0000-0002-2937-1361}{\textcolor[rgb]{0.00,1.00,0.00}{\sf ID}},}$}

%
\address[ECT]{European Centre for Theoretical Studies in Nuclear Physics
            and Related Areas, 
            Villa Tambosi, Strada delle Tabarelle 286, I-38123 Villazzano (TN), Italy}
\address[NJU]{
School of Physics, Nanjing University, Nanjing, Jiangsu 210093, China}
\address[INP]{
Institute for Nonperturbative Physics, Nanjing University, Nanjing, Jiangsu 210093, China\\[1ex]
%
\href{mailto:zyao@ectstar.eu}{zyao@ectstar.eu} (ZQY);
\href{mailto:binosi@ectstar.eu}{binosi@ectstar.eu} (DB);
\href{mailto:cdroberts@nju.edu.cn}{cdroberts@nju.edu.cn} (CDR)
\\[1ex]
Date: 2024 June 13\\[-6ex]
}

\begin{abstract}
Using a symmetry-preserving truncation of the quantum field equations describing hadron properties, parameter-free predictions are delivered for pion and kaon elastic electromagnetic form factors, $F_{P=\pi,K}$, thereby unifying them with kindred results for nucleon elastic electromagnetic form factors.
Regarding positive-charge states, the analysis stresses that the presence of scaling violations in QCD entails that $Q^2 F_P(Q^2)$ should exhibit a single maximum on $Q^2>0$.
Locating such a maximum is both necessary and sufficient to establish the existence of scaling violations.
The study predicts that, for charged $\pi$, $K$ mesons, the $Q^2 F_P(Q^2)$ maximum lies in the neighbourhood $Q^2 \simeq 5\,$GeV$^2$.
Foreseeable experiments will test these predictions and, providing their $Q^2$ reach meets expectations, potentially also provide details on the momentum dependence of meson form factor scaling violation.
\end{abstract}

\begin{keyword}
continuum Schwinger function methods \sep
elastic electromagnetic form factors \sep
emergence of mass \sep
hard exclusive processes \sep
Nambu-Goldstone bosons \sep
scaling violations
\end{keyword}

\end{frontmatter}
\end{CJK}


\section{Introduction}
%
Pions and kaons are Nature's most fundamental Nambu-Goldstone bosons.  They are special because, although being bound-states seeded by a valence quark and valence-antiquark, whose current-masses are identical to those of the proton's valence quarks, the $\pi$ and $K$ are anomalously light hadrons.  For instance, comparing pion and proton masses \cite{Workman:2022ynf}, $m_\pi \approx 0.15 m_p$; in fact, the $\pi$ is just 30\% heavier than the $\mu$-lepton, \emph{i.e}., it has a lepton-like mass.  These features are (\emph{i}) an expression of emergent hadron mass (EHM) within the Standard Model \cite{Horn:2016rip, Eichmann:2016yit, Fischer:2018sdj, Qin:2020rad, Roberts:2021nhw, Binosi:2022djx, Ding:2022ows, Carman:2023zke, Ferreira:2023fva, Deur:2023dzc, Raya:2024ejx, Salme:2022eoy}, owing in part to a set of Goldberger-Treiman relations \cite{Maris:1997tm, Qin:2014vya}, which are exact in QCD; and (\emph{ii}) an accident of Nature, which sees Higgs boson couplings into QCD produce $u$, $d$ quark current masses that are just $4$-$8$-times the electron mass; hence, less than 1\% of $m_p$ \cite{Workman:2022ynf}.  Such contrasts between the $\pi$, $K$, and proton raise many questions, particularly concerning what impressions they leave on the internal structure of these states.

Critically, this is a watershed period because contemporary theory is beginning to deliver robust hadron structure predictions concerning Nambu-Goldstone bosons -- see, \emph{e.g}., Ref.\,\cite{Roberts:2021nhw} and citations thereof; and high-energy, high-luminosity experimental facilities are in operation \cite{Ent:2015kec, Quintans:2022utc}, under construction \cite{Aguilar:2019teb, Arrington:2021biu, AbdulKhalek:2021gbh}, or being planned \cite{Chen:2020ijn, Anderle:2021wcy, Wang:2022xad}.
Thus, after more than one hundred years of accumulating data relating to the structure of the stable proton, such facilities are finally enabling experiment to work with what are effectively $\pi$ and $K$ targets by exploiting Drell-Yan reactions or the Sullivan process \cite{Quintans:2022utc, Aguilar:2019teb, Arrington:2021biu, Chen:2020ijn}.
This will enable collection of the data necessary to test existing pictures of $\pi$, $K$  structure and the new predictions, something previously impossible owing to the short lifetimes of these states.

One of the most striking predictions about $\pi$, $K$ structure, delivered more than 40 years ago, relates to their elastic electromagnetic form factors as measured in a hard exclusive process.  Namely \cite{Lepage:1979zb, Efremov:1979qk, Lepage:1980fj}: for a positive-charge pseudoscalar meson $P$, constituted from valence $f$, $\bar g$ quarks,
{\allowdisplaybreaks
\begin{subequations}
\label{eq:pionFFUV}
\begin{align}
\exists \, Q_0 \gg  m_p & \; |  \;
Q^2 F_P(Q^2) \stackrel{Q^2 > Q_0^2}{\approx} 16 \pi \alpha_s(Q^2)  f_P^2 w_{\varphi_P}^2(Q^2) \,, \\
w_{\varphi_P}^2 & = e_f w_{\varphi_{P^f}}^2 + e_{\bar g} w_{\varphi_{P^{\bar g}}}^2 \\
w_{\varphi_{P^f}} & = \frac{1}{3} \int_0^1 dx\, \frac{1}{x}\, \varphi_P(x;Q^2) \,, \\
w_{\varphi_{P^{\bar g}}} & = \frac{1}{3} \int_0^1 dx\, \frac{1}{1-x}\, \varphi_P(x;Q^2) \,,
\end{align}
\end{subequations}
where
$\alpha_s(Q^2)$ is the QCD running coupling,
$f_P$ is the pseudoscalar meson's leptonic decay constant;
and $\varphi_P(x;Q^2)$ is the meson's leading parton distribution amplitude (DA) evaluated at the hard scale of the interaction.
}

Equation~\eqref{eq:pionFFUV} is remarkable.
It states that beyond some momentum scale, the meson form factor is precisely known:
the magnitude (normalisation) is set by the meson leptonic decay constant, which is an order parameter for dynamical chiral symmetry breaking (DCSB) -- a corollary of EHM;
and scaling violations are apparent, with a $Q^2$ dependence determined by that of the running coupling and evolution of the DA.
On the domain $m_p^2/Q^2 \simeq 0$, $\pi$ and $K$ mesons have the same DA \cite{Lepage:1979zb, Efremov:1979qk, Lepage:1980fj}:
\begin{equation}
\varphi_{\pi,K}(x;Q^2) \to \varphi_{\text{as}}(x) = 6x(1 - x)\,.
\label{phiasy}
\end{equation}
Notwithstanding these things, there are two outstanding issues:
($\mathpzc Q1$) what is the value of $Q_0$ for which scaling violations become apparent;
and
($\mathpzc Q2$) what is the pointwise form of $\varphi_{P}(x;Q^2)$ at the scales for which terrestrial experiments are possible?
These questions can only be answered using a nonperturbative approach to QCD.

The numerical simulation of lattice-regularised QCD (lQCD) provides one nonperturbative scheme for the computation of hadron properties.
Recent studies of light-meson elastic form factors are described, \emph{e.g}., Refs.\,\cite{Chambers:2017tuf, Koponen:2017fvm, Davies:2018zav, Wang:2020nbf, Alexandrou:2021ztx, Ding:2023fac}.  Reviewing such studies, it is evident that existing algorithms and available gauge field configurations do not enable lQCD to answer $\mathpzc Q1$: owing to the level of imprecision and uncertainty in available computations of meson form factors at larger $Q^2$ values, extant lQCD results are insensitive to scaling violations.  (This is illustrated below.)
Regarding $\mathpzc Q2$, on the other hand, useful lQCD results are available for the pointwise behaviour of meson DAs -- see, \emph{e.g}., Ref.\,\cite[Sec.\,8D]{Roberts:2021nhw} and Refs.\,\cite{LatticeParton:2022zqc, Gao:2022vyh}.

Nonperturbative phenomena in QCD have also long been studied using continuum Schwinger function methods (CSMs) \cite{Roberts:2021nhw, Binosi:2022djx, Ding:2022ows, Carman:2023zke, Ferreira:2023fva, Deur:2023dzc, Horn:2016rip, Eichmann:2016yit, Fischer:2018sdj, Qin:2020rad, Raya:2024ejx}.  Following the first studies of $F_\pi(Q^2)$ \cite{Roberts:1994hh, Maris:1998hc}, which explained the expressions of DCSB and the role of symmetries, steady progress has been made.  Today, as described herein, it is possible to deliver predictions that unify $F_{\pi,K}(Q^2)$ with nucleon elastic electromagnetic form factors and address both $\mathpzc Q1$, $\mathpzc Q2$.

\section{Meson Form Factors: Current Status}
Given the verification challenges presented to experiment and theory by Eq.\,\eqref{eq:pionFFUV} and the simplicity of what may seem, from a quantum mechanics perspective, to be merely a two-body problem, $\pi$ and $K$ elastic electromagnetic form factors have been a longstanding focus of QCD phenomenology and theory.  This interest has been amplified by the promises contained in a new era of $\pi$ and $K$ targets, so that many new studies have been completed -- see, \emph{e.g}., Refs.\,\cite{Ahmady:2018muv, Choi:2020xsr, Lan:2021wok, Ydrefors:2021dwa, Er:2022cxx, Li:2022mlg, Ahmed:2023zkk}.

\begin{figure}[t]
\centerline{%
\includegraphics[clip, width=0.42\textwidth]{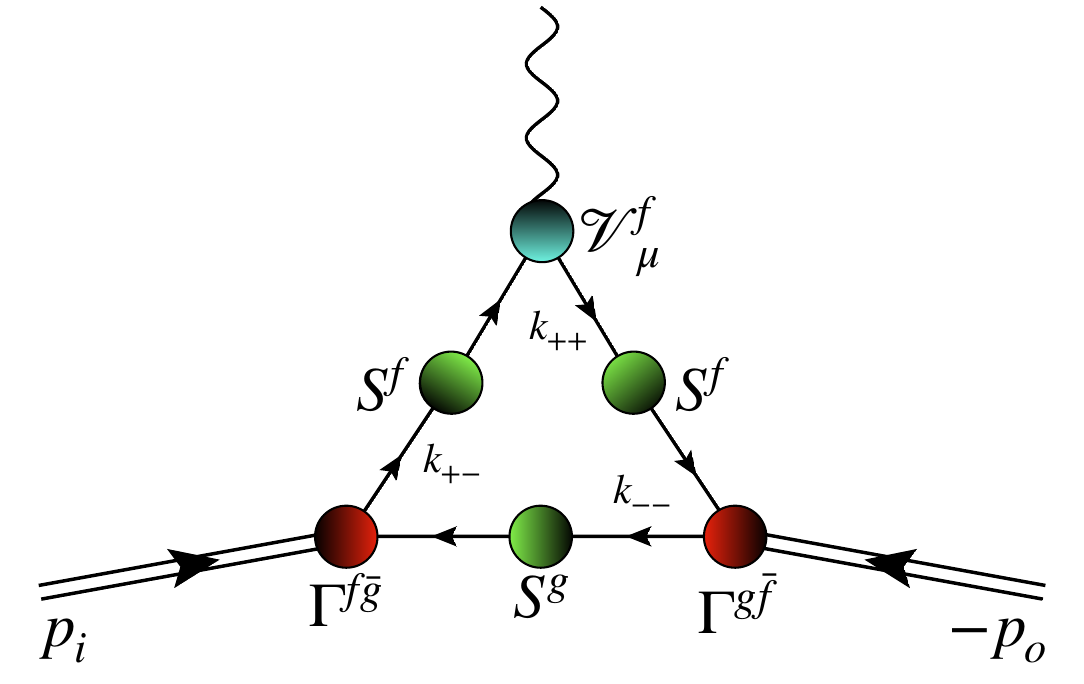}}
\caption{\label{FigME}
Matrix element characterising the $f \bar g$ meson elastic electromagnetic form factor in rainbow-ladder (RL) truncation.
Legend.
$\Gamma^{f\bar g}$ -- Bethe-Salpeter amplitude for the $f \bar g$ meson;
$S^{f, g}$ -- dressed quark propagators;
${\mathpzc V}_\mu^{f,\bar g}$ -- dressed photon--$f,\bar g$-quark vertex.
All quantities calculated in RL truncation.
}
\end{figure}

The first CSM analysis to present a unified treatment of $F_{\pi,K}(Q^2)$ is reported in Ref.\,\cite{Maris:2000sk}.  It calculated the form factors at leading order in the systematic, symmetry-preserving CSM truncation scheme introduced in Refs.\,\cite{Munczek:1994zz, Bender:1996bb}, \emph{i.e}., rainbow-ladder (RL) truncation, at which level the form factors are obtained from the matrix element illustrated in Fig.\,\ref{FigME}.

\begin{figure}[t]
\vspace*{1.5em}

\includegraphics[width=0.41\textwidth]{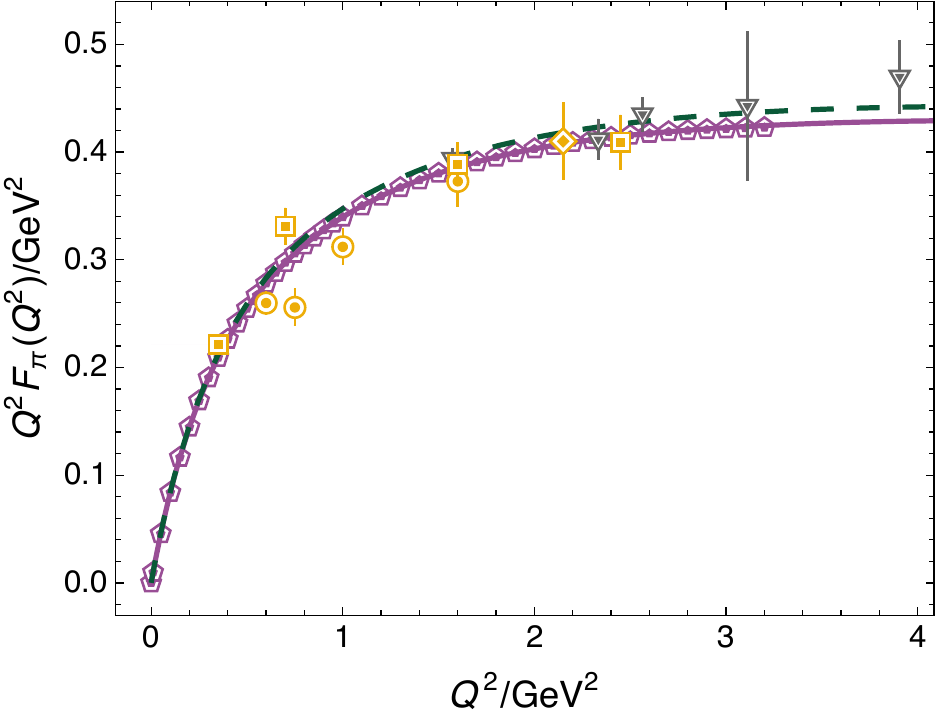}
\vspace*{-40ex}

\leftline{\hspace*{0.1em}{\large{\textsf{A}}}}

\vspace*{40ex}

\includegraphics[width=0.41\textwidth]{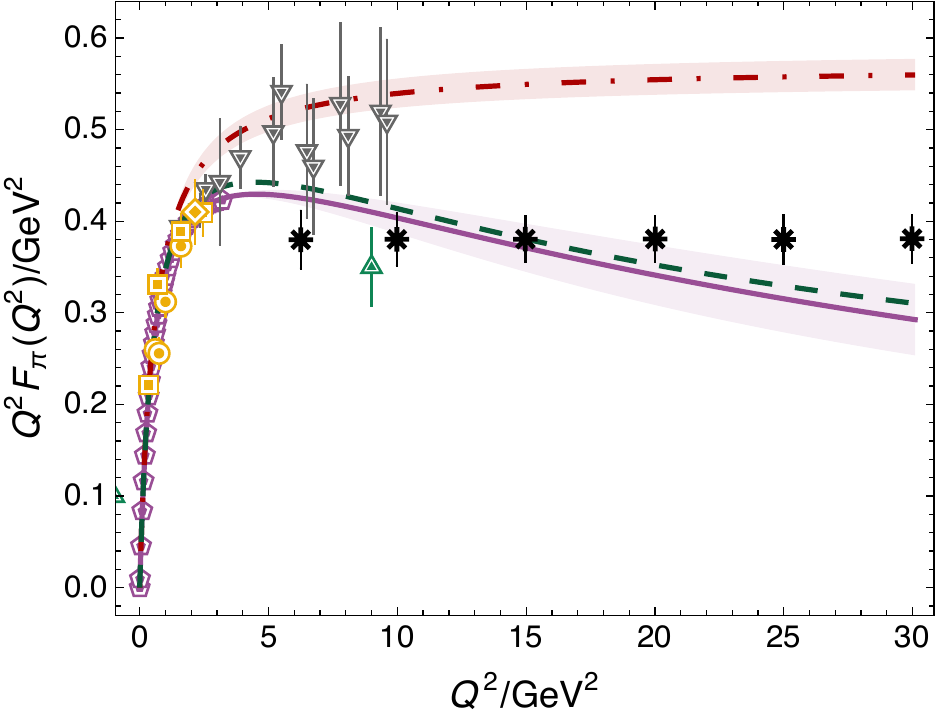}
\vspace*{-40ex}

\leftline{\hspace*{0.1em}{\large{\textsf{B}}}}

\vspace*{36ex}

\caption{\label{FigFpi}
Pion elastic electromagnetic form factor, $Q^2 F_\pi(Q^2)$.
%
Legend.
Purple curves -- SPM interpolations [Sec.\,\ref{SecSPM}] of results obtained herein (purple pentagons) and their extrapolation onto $Q^2 > Q_{\pi^m}^2=3.2\,$GeV$^2$;
dashed green curve  -- CSM results obtained using PTIRs \cite{Chang:2013nia};
grey down-triangles -- lQCD \cite{Ding:2023fac}.
Data (gold) -- diamond \cite{Horn:2007ug}; circles and squares \cite{Huber:2008id}.
Panel B only.
Dot-dashed red curve -- monopole with mass fixed by empirical pion charge radius, $r_\pi \approx 0.66\,$fm.
Further: green up triangle -- anticipated uncertainty of forthcoming JLab measurement at the highest accessible $Q^2$ point \cite{E12-19-006};
and black asterisks -- expected uncertainty of EIC data, whose coverage should extend to $Q^2 \approx 35\,$GeV$^2$ \cite{Aguilar:2019teb, Arrington:2021biu}.
The central magnitude of these points was chosen arbitrarily.
}
\end{figure}

\begin{figure}[t]
\vspace*{1.5em}

\includegraphics[width=0.41\textwidth]{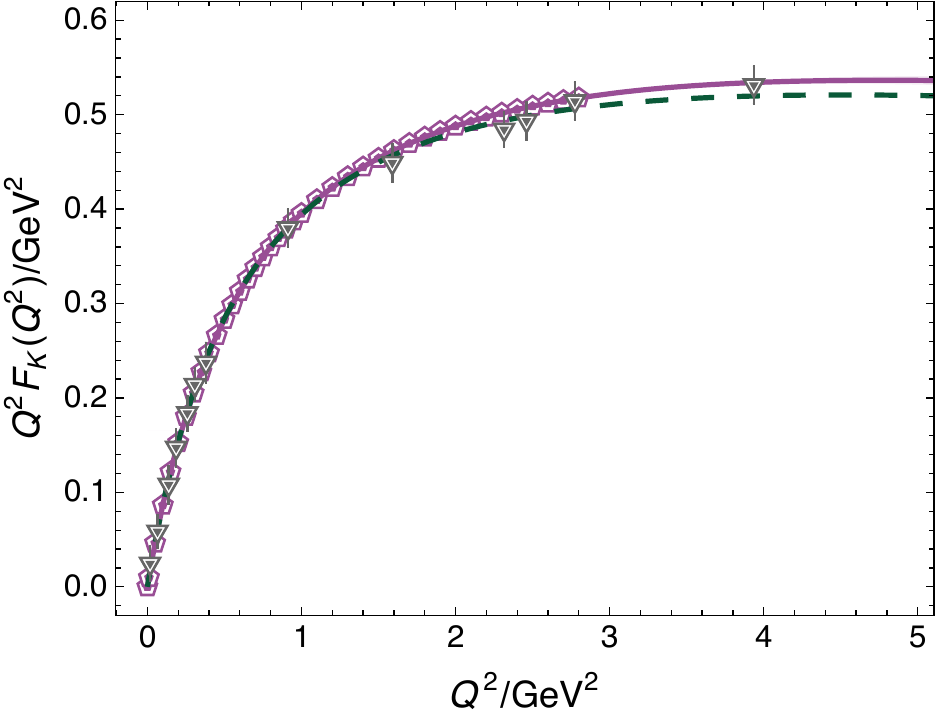}
\vspace*{-40ex}

\leftline{\hspace*{0.1em}{\large{\textsf{A}}}}

\vspace*{40ex}

\includegraphics[width=0.41\textwidth]{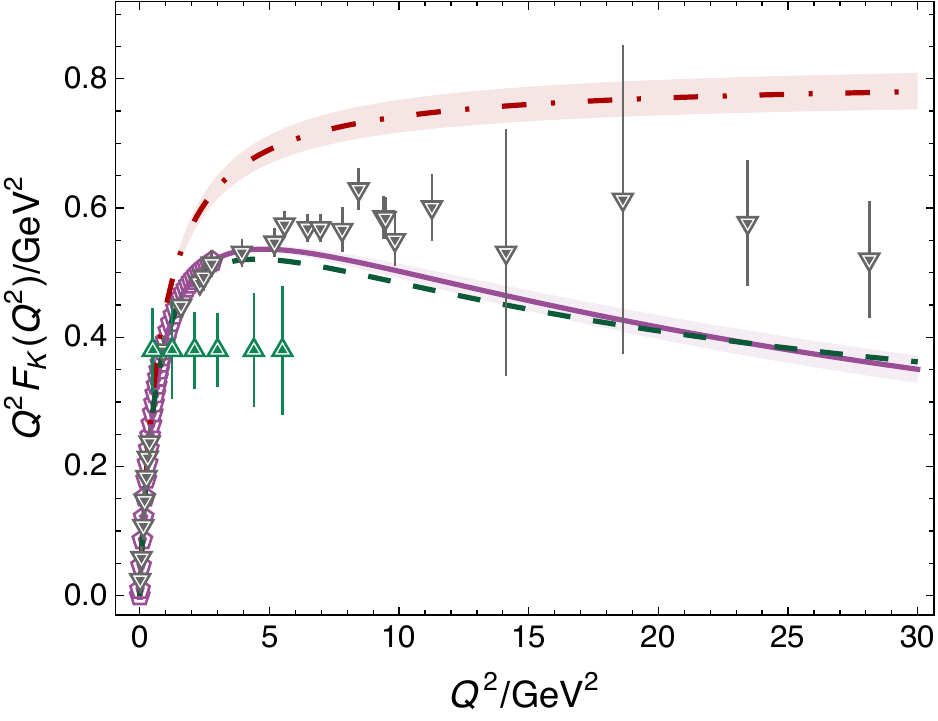}
\vspace*{-40ex}

\leftline{\hspace*{0.1em}{\large{\textsf{B}}}}

\vspace*{36ex}

\caption{\label{FigFKp}
Charged kaon elastic electromagnetic form factor, $Q^2 F_{K}(Q^2)$.
%
Legend.
Purple curves -- SPM interpolations of results obtained herein (purple pentagons) and their extrapolation onto $Q^2 > Q_{K^m}^2=4.9\,$GeV$^2$;
dashed green curve  -- CSM results obtained using PTIRs \cite{Gao:2017mmp};
grey down-triangles -- lQCD \cite{Ding:2023fac}.
At present, there are no precise charged kaon form factor data.
Panel B only.
Dot-dashed red curve -- monopole with mass fixed by empirical kaon charge radius, $r_K \approx 0.54\,$fm \cite[SPM]{Cui:2021aee}.
Further: green up-triangles illustrate the expected coverage and precision of anticipated JLab data \cite{E12-09-011}.  The central magnitude of these points was chosen arbitrarily.}
\end{figure}

The numerical algorithms used in Ref.\,\cite{Maris:2000sk} precluded access to the domain $Q^2 \gtrsim 4\,$GeV$^2$.  (This point is discussed further below.)  Such challenges were overcome in Refs.\,\cite{Chang:2013nia, Gao:2017mmp} by using perturbation theory integral representations (PTIRs) \cite{Nakanishi:1969ph} of all Schwinger functions involved in calculating the relevant matrix elements -- see Fig.\,\ref{FigME}, with the results displayed as dashed green curves in Figs.\,\ref{FigFpi}, \ref{FigFKp}.
Regarding the pion, the PTIR-based prediction indicates that scaling violations will become apparent on $Q^2 \gtrsim 5\,$GeV$^2$.  Given the errors anticipated in large-$Q^2$ pion form factor experiments at Jefferson Laboratory \cite{E12-19-006, E12-09-011}, those measurement may be able to validate this prediction.  It is certain that the prediction can be tested at the electron ion collider (EIC) \cite[Fig.\,4.12B]{Roberts:2021nhw}.
Similar statements hold for the charged-kaon form factor \cite[Fig.\,4.14A]{Roberts:2021nhw}.

\section{Unification of Nucleon and Meson Form Factors}
Hitherto, no single framework or study has succeeded in delivering a unified set of predictions for pion, kaon and nucleon elastic electromagnetic form factors.  We remedy that herein by extending the three-body Faddeev equation analysis of nucleon electromagnetic form factors in Ref.\,\cite{Yao:2024uej} to $F_{\pi,K}(Q^2)$.

The starting point for this completion is the matrix element in Fig.\,\ref{FigME}, which describes the photon coupling to the meson's valence $f$ quark:
\begin{align}
P_\mu F_P^f(Q^2) & = {\rm tr}_{\rm CD} \int \frac{d^4 k}{(2\pi)^4}
S^f(k_{++})  {\mathpzc V}_\mu^f (k_{++},k_{+-}) S^f(k_{+-}) \nonumber \\
&
\times  \Gamma_P^{f\bar g}(k_{+-},k_{--};p_i)
S^g(k_{--}) \Gamma_P^{g \bar f}(k_{--},k_{++};-p_o)\,,
\label{EqME}
\end{align}
where the trace is over colour and spinor indices,
$p_{o,i}= P \pm Q/2$, $p_{o,i}^2 = -m_P^2$, $P\cdot Q = 0$,
$k_{\pm\pm} = k \pm P/2 \pm Q/2$,
and each element in the integrand is calculated using RL truncation.  The complete meson form factor is
\begin{equation}
F_P (Q^2) = e_f F_P^f(Q^2) + e_{\bar g} F_P^{\bar g}(Q^2)\,,
\label{EqFPQ2}
\end{equation}
where $e_f$ is the charge of the struck quark in units of the positron charge and $F_P^{\bar g}(Q^2)$ is plain by analogy with Eq.\,\eqref{EqME}.

Discussions of the formulation and solution of the RL truncation of the Dyson-Schwinger equation for each element in Eq.\,\eqref{EqME} can be found in, \emph{e.g}., Refs.\,\cite{Maris:1997tm, Maris:2000sk}.  In all such analyses, the key is the quark+quark scattering kernel, for which the RL truncation is obtained by writing \cite{Maris:1997tm}:
{\allowdisplaybreaks
\begin{subequations}
\label{EqRLInteraction}
\begin{align}
\label{KDinteraction}
\mathscr{K}_{tu}^{rs}(k) & = {\mathpzc G}_{\mu\nu}(k) [i\gamma_\mu\frac{\lambda^{a}}{2} ]_{ts} [i\gamma_\nu\frac{\lambda^{a}}{2} ]_{ru}\,,\\
 {\mathpzc G}_{\mu\nu}(k)  & = \tilde{\mathpzc G}(y) T_{\mu\nu}(k)\,,
\end{align}
\end{subequations}
$k^2 T_{\mu\nu}(k) = k^2 \delta_{\mu\nu} - k_\mu k_\nu$,  $y=k^2$.  The tensor structure specifies Landau gauge, used because it is a fixed point of the renormalisation group and that gauge for which corrections to RL truncation are minimised \cite{Bashir:2009fv}.
In Eq.\,\eqref{EqRLInteraction}, $r,s,t,u$ represent colour and spinor matrix indices (as necessary).
}

A realistic form of ${\mathpzc G}_{\mu\nu}(y)$ is explained elsewhere \cite{Qin:2011dd, Binosi:2014aea}:
\begin{align}
\label{defcalG}
 \tilde{\mathpzc G}(y) & =
 \frac{8\pi^2}{\omega^4} D e^{-y/\omega^2} + \frac{8\pi^2 \gamma_m \mathcal{F}(y)}{\ln\big[ \tau+(1+y/\Lambda_{\rm QCD}^2)^2 \big]}\,,
\end{align}
where $\gamma_m=12/25$, $\Lambda_{\rm QCD} = 0.234\,$GeV, $\tau={\rm e}^2-1$, and ${\cal F}(y) = \{1 - \exp(-y/\Lambda_{\mathpzc I}^2)\}/y$, $\Lambda_{\mathpzc I}=1\,$GeV.
%
%
We employ a mass-independent (chiral-limit) momentum-subtraction renormalisation scheme \cite{Chang:2008ec}. 

Following the nucleon study \cite{Yao:2024uej}, we set $\omega = 0.8\,$GeV.  Then, with $\omega D = 0.8\,{\rm GeV}^3$ and renormalisation point invariant light-quark current mass $\hat m_u = \hat m_d = 6.04\,$MeV, which corresponds to a one-loop mass at $\zeta_2=2\,$GeV of $4.19\,$MeV, one obtains: pion mass $m_\pi = 0.14\,$GeV; pion leptonic decay constant $f_\pi=0.094\,$GeV; and $m_p=0.94\,$GeV.  (We assume isospin symmetry throughout.)
Introducing the $s$ quark, with $\hat m_s = 0.117\,$GeV $\Rightarrow m_s^{\zeta_2} = 0.081\,$GeV, then one arrives at calculated values $m_K=0.45\,$GeV, $f_K = 0.11\,$GeV.
%
Completing the picture, here we list calculated masses and leptonic decay constants for the $\rho$, $K^\ast$, $\phi$ mesons (in GeV):
\begin{equation}
\begin{array}{l|llllll}
& m_\rho & m_{K^\ast} & m_\phi & f_\rho & f_{K^\ast} & f_\phi \\\hline
{\rm herein} & 0.72\ & 0.93\ & 1.06\ & 0.14\ & 0.17\ & 0.19 \\
\mbox{\cite[PDG]{Workman:2022ynf}} & 0.78\ & 0.89\ & 1.02\ & 0.15\ & 0.16\ & 0.17\
\end{array} .
\end{equation}
Compared with empirical values \cite{Workman:2022ynf}, the calculated results deliver a mean absolute relative difference of $4(3)$\%.
%
Notably, so long as the product $\omega D$ is kept fixed, then physical observables remain practically unchanged under $\omega \to (1\pm 0.2)\omega$ \cite{Qin:2020rad}.

The interaction involves one parameter and there are two quark current-masses.  These quantities are now fixed.  Hence, hereafter, all calculations are parameter-free.

It is worth remarking that, following Ref.\,\cite{Qin:2011dd}, one may draw a connection between the interaction in Eq.\,\eqref{defcalG} 
and QCD's process-independent running coupling, discussed in Refs.\,\cite{Cui:2019dwv, Deur:2023dzc}.
That coupling is characterised by an infrared value $\hat\alpha(0)/\pi = 0.97(4)$ and a gluon mass $\hat m_0 = 0.43(1)\,$GeV.
The analogous quantities inferred from Eq.\,\eqref{defcalG} are: 
\begin{equation}
\label{DiscussInteraction}
\alpha_{\mathpzc G}(0)/\pi = 1.45\,,\quad m_{\mathpzc G} = 0.54\,{\rm GeV}\,.
\end{equation}
Evidently, contemporary formulations of the RL truncation draw a sound bridge to QCD, especially when one recalls that earlier implementations yielded $\alpha_{\mathpzc G}(0)/\pi \approx 15$, \emph{i.e}., an infrared value ten-times larger \cite{Qin:2011dd}.

Numerical methods for solving sets of coupled gap and Bethe-Salpeter equations are described, \emph{e.g}., in Refs.\,\cite{Maris:1997tm, Maris:2005tt, Krassnigg:2009gd}.  These schemes were used to obtain the meson results discussed above and, hereafter, to solve all equations relevant to calculation of the matrix element in Eq.\,\eqref{EqME} so as to arrive at predictions for pion and kaon form factors.  Kindred predictions for all nucleon elastic form factors are described in Ref.\,\cite{Yao:2024uej}.

\section{Covering large $Q^2$}
\label{SecSPM}
Before proceeding, a technical remark is appropriate.
When working with an interaction in the class which contains Eq.\,\eqref{defcalG}, one typically finds that the dressed-quark propagators possess complex conjugate poles \cite{Maris:1997tm}.  In the evaluation of form factors, these poles introduce singularities that move into the complex-$k^2$ domain sampled by the bound-state equations \cite{Bhagwat:2002tx}.  Consequently, for each meson, there is a maximum value of $Q^2=Q_{P^m}^2$ beyond which evaluation of the integrals required by Eq.\,\eqref{EqFPQ2} is no longer possible with conventional algorithms.

As noted above, Refs.\,\cite{Chang:2013nia, Gao:2017mmp} circumvented this moving singularity problem by using PTIRs for each matrix-valued function in the integrand.
Constructing accurate PTIRs is, however, time consuming.
In our case, PTIR interpolations would need to be developed for $34$ scalar functions.
Considering nucleon electromagnetic form factors, the number is $128$.

We therefore choose to proceed along another route, which was recently opened following rediscovery and improvement of a multipoint Pad\'e approximant approach for the interpolation and extrapolation of analytic functions.  Now often called the Schlessinger point method (SPM), its properties and accuracy are explained and illustrated, \emph{e.g}., in Refs.\,\cite{Schlessinger:1966zz, PhysRev.167.1411, Tripolt:2016cya, Chen:2018nsg, Binosi:2018rht, Binosi:2019ecz, Huber:2020ngt, Fischer:2020xnb, Yao:2021pyf, Cui:2021gzg, Cui:2021aee, Cui:2022fyr, Binosi:2022ydc, Cui:2022dcm}.  We note only that the SPM is able to reliably reconstruct a function in the complex plane within a radius of convergence determined by that one of the function's branch points located nearest to the real domain from which the sample points are drawn.  Modern implementations, furthermore, introduce a statistical element; consequently, all extrapolations come with an objective and reliable estimate of uncertainty.  Thus, as done recently for nucleon form factors \cite{Yao:2024uej}, we use the SPM to obtain predictions for all form factors on the domain $Q^2 > Q_{P^m}^2$.

Working with the flavour-separated form factors defined by Eq.\,\eqref{EqME}, we develop SPM interpolations and extrapolations as follows.  (\emph{N.B}.  In the isospin symmetry limit, $F_\pi^u \equiv F_\pi$.)
\begin{description}
\item[Step 1] For $F_\pi^u$, we produce $N=44$ directly calculated values of $Q^2 F_\pi^u(Q^2)$, spaced evenly on $Q^2 \leq Q_{P^m}^2$; and for $F_K^u$, $F_K^{\bar s}$, $N=57$ such values.

\item [Step 2]  From that set, $M_0=8$ points are chosen at random,
the usual SPM continued fraction interpolation is constructed,
and that function is extrapolated onto $Q^2 > Q_{P^m}^2$.

Regarding $Q^2 F_\pi^u(Q^2)$, the curve is retained so long as it is singularity free and concave on $Q^2\leq 100\,$GeV$^2$.
Notably, we do not require that the curve be positive definite.
Nevertheless, all randomly generated SPM curves extrapolate with this feature.

Concerning the kaon,
we applied an additional constraint; namely, that the $Q^2=100\,$GeV$^2$ value of the extrapolation of the ratio $F_K^{\bar s}/F_K^u$, formed from the individual interpolators, is less than its value at $Q_{K^m}^2$.  (This is a rudimentary way to comply with an ultraviolet consequence of Eq.\,\eqref{eq:pionFFUV}, \emph{viz}.\ $F_K^{\bar s}(Q^2)/F_K^{u}(Q^2) \simeq 1$ on $m_p^2/Q^2 \simeq 0$.)

\item[Step 3]
Step 2 is repeated with another set of $M_0$ randomly chosen points.
Combined, Steps 2 and 3 admit $\approx 2\times 10^{10}$ independent extrapolations for $Q^2F_\pi^u(Q^2)$
and
$\approx 7 \times 10^{11}$ for both $Q^2F_K^{u,\bar s}(Q^2)$.

\item[Step 4]
One continues with 2 and 3 until $n_{M_0}=500\,$ smooth extrapolations are obtained.

\item[Step 5]
Steps 2 and 3 are repeated for $M=\{M_0+ 2 i | i=1,\ldots,6\}$.
Only interpolants obtained with an even number of input points are chosen because then the continued fraction vanishes as $1/Q^2$ in the far ultraviolet.  At any finite $Q^2$, scaling violations are produced by interference between distinct power terms in the numerator and denominator.

\item [Step 6]
At this point, one has $3\,500$ statistically independent extrapolations for $Q^2 F_\pi(Q^2)$ and $Q^2F_K^{u,\bar s}(Q^2)$.

\end{description}
Working with this set of extrapolations, the pion, charged and neutral kaon form factors are obtained using Eq.\,\eqref{EqFPQ2}.  At each value of $Q^2$ for each form factor, we record the mean value of all curves as the central prediction and, as the uncertainty, report the function range which contains 68\% of all extrapolations -- this is a $1\sigma$ band.

\begin{table}[t]
\caption{\label{TabRadii}
Predictions for $\pi$ and $K$ charge radii compared with values inferred from experiment \cite[PDG]{Workman:2022ynf}, \cite[SPM]{Cui:2021aee}.
As stressed elsewhere \cite[Sec.\,6]{Cui:2022fyr}, the precision of available data is insufficient to enable an objective empirical determination of $r_{K^+}$.
(Dimensioned quantities in fm$^2$.)}
\begin{center}
\begin{tabular*}
{\hsize}
{
l@{\extracolsep{0ptplus1fil}}
|l@{\extracolsep{0ptplus1fil}}
l@{\extracolsep{0ptplus1fil}}
l@{\extracolsep{0ptplus1fil}}}\hline\hline
 & herein & Exp. & SPM  \\\hline
 $r_\pi^2\ $ & $\phantom{-}[0.67]^2\ $ & $\phantom{-}[0.663(6)]^2 $ & $[0.640(7)]^2\ $ \\
 $r_{K^\pm}^2\ $ & $\phantom{-}[0.60]^2\ $ & $\phantom{-}[0.56(3)]^2 $ & $[0.536(6)]^2\ $ \\
 $r_{K^0}^2\ $ & $-[0.30]^2\ $ & $-[0.277(18)]^2 $ &
 \\\hline\hline
\end{tabular*}
\end{center}
\end{table}

\section{Pion Form Factor}
Our prediction for $Q^2 F_\pi(Q^2)$ is drawn in Fig.\,\ref{FigFpi}.  Defined in the usual way, the charge radius -- see Table~\ref{TabRadii} -- is in agreement with experiment.  Furthermore, the $Q^2$-dependence of the result matches that suggested by the most recent precision measurements \cite{Horn:2007ug, Huber:2008id} -- see Fig.\,\ref{FigFpi}A.
The small mismatch on $Q^2\lesssim 4\,$GeV$^2$ between our result and that in Ref.\,\cite{Chang:2013nia} owes to modest differences between our $(\omega,D)$ values and those used therein.  Our choice is informed by contemporary developments in understanding QCD's process-independent effective charge -- see Eq.\,\eqref{DiscussInteraction}, information that was not available to Ref.\,\cite{Chang:2013nia}.

The larger-$Q^2$ CSM prediction, obtained using the SPM, is displayed in Fig.\,\ref{FigFpi}B.  Combining both panels in Fig.\,\ref{FigFpi}, one sees good agreement between our prediction and that obtained using PTIRs \cite{Chang:2013nia}.  This increases confidence in both results.

In Fig.\,\ref{FigFpi}B, we have also drawn a single-pole vector meson dominance (VMD) estimate for $Q^2 F_\pi(Q^2)$, with the pole mass chosen to reproduce $r_\pi$.  This curve begins to deviate significantly from data at the upper bound of available measurements.  Notably, however, the precision of existing lQCD results is insufficient to distinguish between the VMD estimate and the CSM prediction.

A salient feature of the CSM predictions is the existence of a maximum in $Q^2 F_\pi(Q^2)$, which occurs on $Q^2\simeq 4.6(5)\,$GeV$^2$.  Thereafter, scaling violation is apparent, with $Q^2 F_\pi(Q^2)$ falling steadily toward zero as $\approx [1/\ln Q^2]^{\gamma_F}$, $\gamma_F \approx 1.1$ on the displayed domain, consistent with Eq.\,\eqref{eq:pionFFUV}.
Any model that expresses VMD must exhibit the opposite trend, \emph{viz}.\ a result for $Q^2 F_\pi(Q^2)$ which rises steadily with increasing $Q^2$, without inflection, toward its finite ultraviolet limit.


Figure~\ref{FigFpi}B also displays points with the anticipated precision of new era data to be collected at Jefferson Laboratory (JLab) and the electron ion collider (EIC).
Based on these projections, JLab has the potential to discover scaling violation in a hard exclusive process; but failing that, it should be found at EIC or any other facility with similar $Q^2$ reach and precision.

\section{Flavour Separation of Meson Form Factors}
Working with Eqs.\,\eqref{EqME}, \eqref{EqFPQ2}, the primary quantities in the calculation of meson form factors are the individual valence quark contributions.  As noted above, owing to isospin symmetry, $F_\pi^u = F_\pi$; moreover, $F_{K^0}^d = F_K^u$.  (\emph{N.B}.\ Here and hereafter, the quark electric charges are usually factored out.)

\begin{figure}[t]
\vspace*{1.5em}

\includegraphics[width=0.41\textwidth]{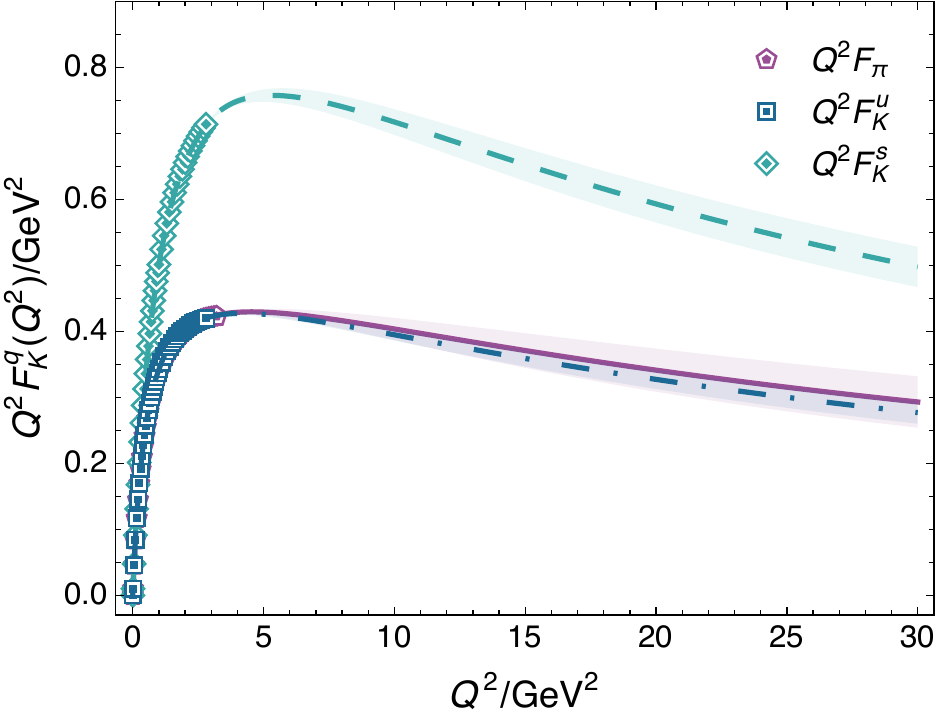}
\vspace*{-40ex}

\leftline{\hspace*{0.1em}{\large{\textsf{A}}}}

\vspace*{40ex}

\includegraphics[width=0.41\textwidth]{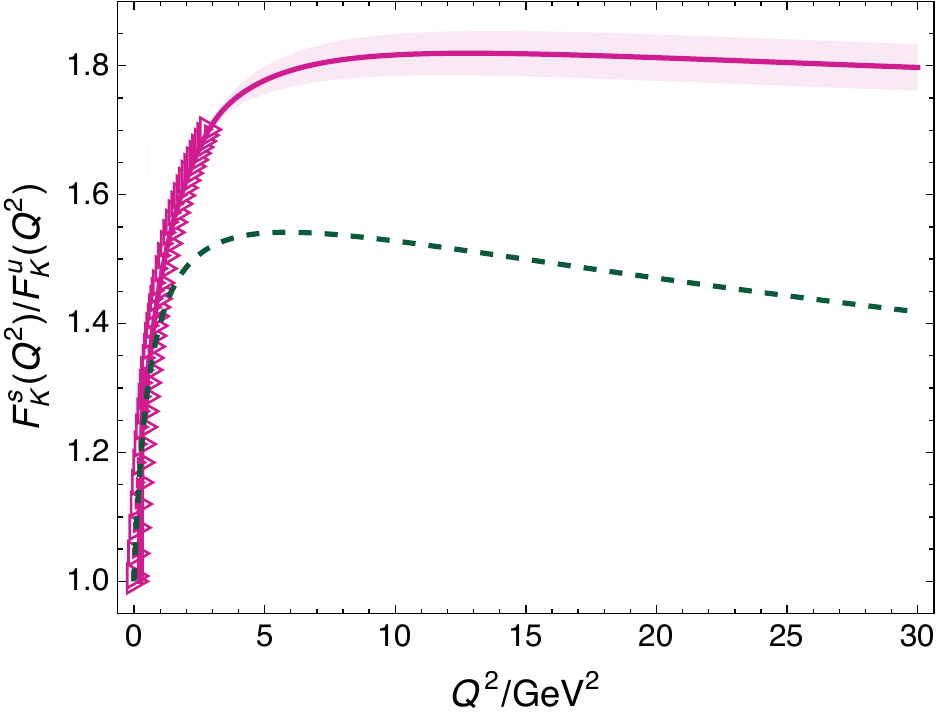}
\vspace*{-40ex}

\leftline{\hspace*{0.1em}{\large{\textsf{B}}}}

\vspace*{36ex}

\caption{\label{FigFS}
{\sf Panel A}.
Flavour separation of pseudoscalar meson form factors, obtained using Eqs.\,\eqref{EqME}, \eqref{EqFPQ2}, and their associated extrapolations (the extent of direct calculation is evident from the plotted symbols):
$F_K^{\bar s}$ -- cyan dashed curve;
$F_K^u$ -- blue dot-dashed curve;
$F_\pi^u$ -- purple solid curve.
{\sf Panel B}. $F_K^{\bar s}/F_K^{u}$: magenta solid curve -- herein; and dashed green curve -- PTIR \cite{Gao:2017mmp}.
}
\end{figure}

Our predictions for all independent valence-quark form factor contributions are displayed in Fig.\,\ref{FigFS}A.
Consistent with Ref.\,\cite{Maris:2000sk}, $F_K^u\approx F_\pi^u$ on the entire $Q^2$ domain, \emph{viz}.\ insofar as light quarks are concerned, there is little sensitivity to their host environment.
On the other hand, $F_K^{\bar s} > F_K^{u}$ $\forall Q^2 > 0$.
Moreover, in agreement with Refs.\,\cite{Chang:2013nia, Gao:2017mmp} and consistent with Eq.\,\eqref{eq:pionFFUV}, each valence-quark form factor is positive definite on $Q^2\geq 0$.

The $\bar s/u$ ratio in the kaon is drawn in Fig.\,\ref{FigFS}B.
Owing to current conservation, this ratio must be unity at $Q^2=0$.
It rises with increasing $Q^2$ because the $\bar s$ quark is heavier than the $u$ quark; hence, as seen in Fig.\,\ref{FigFS}A, the $\bar s$-in-$K$ form factor falls more slowly with $Q^2$ (it is associated with a smaller radius).
Now, returning to Eq.\,\eqref{eq:pionFFUV}, it is plain that, in QCD, this ratio must be approximately unity on $m_p^2/Q^2 \simeq 0$; so, it exhibits a single maximum at spacelike $Q^2$.  (Pseudoscalar meson form factors are monotonically decreasing functions.)
Again, the appearance of that maximum is a signal for the onset of scaling violation in the kaon form factor and the height of the peak is an expression of Higgs boson modulation of EHM.
We predict that the maximum lies on $Q^2\simeq 5.9 (2.6)\,$GeV$^2$ with peak height $\bar s/u = 1.76(8)$.  Evidently, these values are qualitatively and semi-quantitatively consistent with the PTIR results \cite{Gao:2017mmp}.

\section{Charged Kaon Form Factor}
The charged kaon form factor,
$F_K = e_u F_K^u + e_{\bar s} F_K^{\bar s}$,
is drawn in Fig.\,\ref{FigFKp}.
As precise $F_{K}(Q^2)$ data are lacking, no comparison with experiment is made in these panels.  The predicted charge radius is commensurate with available empirical estimates -- see Table~\ref{TabRadii}.
Notably, since $m_K^2$ is an order of magnitude larger than $m_\pi^2$, then $Q_{K^m}^2 = 4.9\,$GeV$^2>Q_{\pi^m}^2$.

The larger-$Q^2$ SPM-based CSM prediction is depicted in Fig.\,\ref{FigFKp}B.  Combining both panels in Fig.\,\ref{FigFKp}, good agreement between our prediction and the PTIR result \cite{Gao:2017mmp} is apparent, increasing confidence in both.  Thus, one now has a clear indication of the magnitude and evolution of $F_K$: both studies predict a maximum in $Q^2 F_{K}(Q^2)$ on $Q^2\simeq 4.8(5)\,$GeV$^2$ and, subsequently, scaling violation, with a steady $[1/\ln Q^2]^{\gamma_F}$ decrease toward zero.

Figure~\ref{FigFKp}B also displays a single-pole VMD estimate for $Q^2 F_K(Q^2)$, with the pole mass chosen to reproduce $r_K=0.54(1)\,$fm.  Both CSM analyses predict a discernible departure from the VMD curve at practically the same location.
Comparing $F_{\pi,K}$, it is notable that, with increasing mass of the meson involved, the position at which the VMD curve separates from the direct calculation moves to smaller $Q^2$ and the difference between the curves increases.  This is a general feature \cite{Chen:2018rwz}.
Considering their anticipated $Q^2$ reach, JLab experiments \cite{E12-09-011} may be expected to confirm this breakaway.
On the other hand, discovery of scaling violation in the charged kaon form factor may require a machine with EIC capabilities.
Further in this connection, it is evident from Fig.\,\ref{FigFKp}B that existing lQCD results are insensitive to scaling violation in $F_K(Q^2)$: on $Q^2 \gtrsim 5\,$GeV$^2$, the results are consistent with $Q^2F_K(Q^2)=\,$constant.

\begin{figure}[t]
\includegraphics[width=0.41\textwidth]{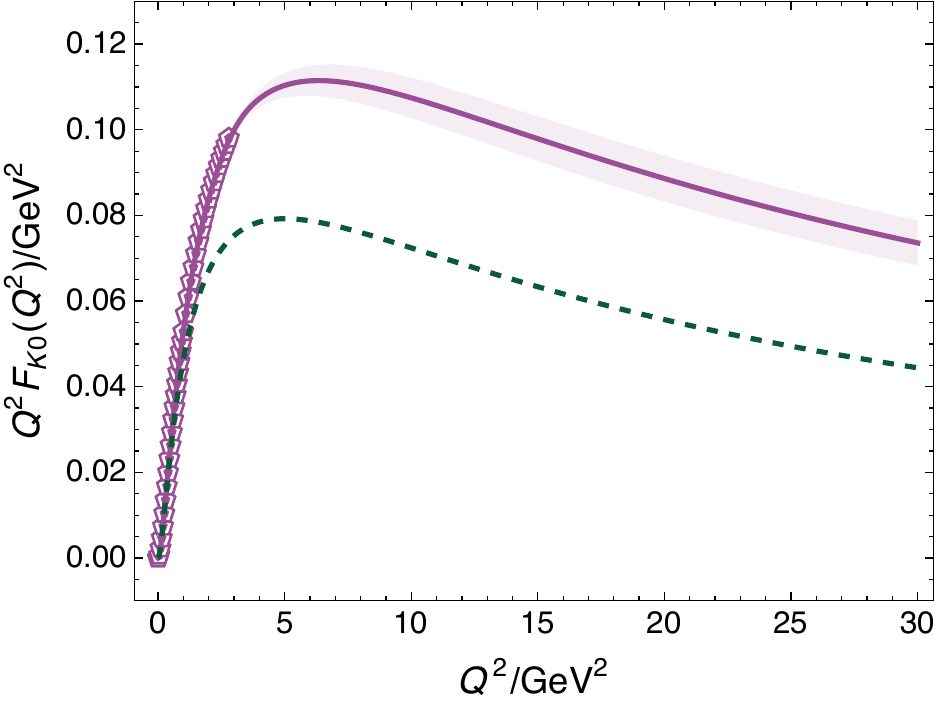}

\caption{\label{FigFK0}
Neutral kaon elastic electromagnetic form factor, $Q^2 F_{K^0}(Q^2)$.
%
Legend.
Purple curves -- SPM interpolations of results obtained herein (purple pentagons) and their extrapolation onto $Q^2 > 4.9\,$GeV$^2$;
dashed green curve  -- CSM results obtained using PTIRs \cite{Gao:2017mmp}.
At present, there are no precise neutral kaon data.
}
\end{figure}

\section{Neutral Kaon Form Factor}
The neutral kaon form factor,
$F_K = e_d F_K^d + e_{\bar s} F_K^{\bar s}$,
is drawn in Fig.\,\ref{FigFK0}.
This difference is a keen measure of Higgs boson modulation of EHM, a fact highlighted by the quantitative difference between our prediction and that obtained using PTIRs \cite{Gao:2017mmp}.  A minor alteration in the description of $\bar s$-in-$K$ physics when developing PTIR approximations or SPM extrapolations, as viewed from the charged-kaon perspective, translates into a noticeable change in results for $F_{K^0}$.

It is worth noting that the curves in Fig.\,\ref{FigFK0} are consistent with Figs.\,\ref{FigFpi}\,--\,\ref{FigFS}.
Using environment insensitivity, one reads from Fig.\,\ref{FigFpi} that both our study and Ref.\,\cite{Gao:2017mmp} have $F_K^{u} \approx F_\pi^u$.
Figure~\ref{FigFKp} then shows that $F_K^{\bar s}$ herein is larger than the same function obtained in Ref.\,\cite{Gao:2017mmp}; consequently, we predict a somewhat larger difference $F_K^{\bar s}-F_K^{d}$.

Notably, if choosing to combine the extrapolation results instead of contrasting them, then one may conclude from these analyses that the large $Q^2$ behaviour of the neutral kaon form factor is known with an uncertainty at the level of $\approx 20$\%.

\section{Meson Distribution Amplitudes}
If the asymptotic DA, Eq.\,\eqref{phiasy}, is used in Eq.\,\eqref{eq:pionFFUV}, along with the one-loop QCD running coupling, and empirical values for the leptonic decay constants, then
{\allowdisplaybreaks
\begin{subequations}
\label{FPasy}
\begin{align}
\left. Q^2 F_\pi(Q^2)\right|_{Q^2 = 15\,{\rm GeV}^2}^{\varphi_{\rm as}} & = 0.12\,{\rm GeV}^2, \\
\left. Q^2 F_K(Q^2)\right|_{Q^2 = 15\,{\rm GeV}^2}^{\varphi_{\rm as}} & = 0.16\,{\rm GeV}^2.
\end{align}
\end{subequations}
Reviewing Figs.\,\ref{FigFpi}, \ref{FigFKp}, these values are roughly three-times smaller than our predictions.  Within reasonable bounds, the outcome is practically independent of the value of $Q^2$ chosen for the comparison.  The mismatch has long been recognised and Refs.\,\cite{Chang:2013nia, Gao:2017mmp} pointed to (part of) a possible solution.  Namely, one should not make this comparison using $\varphi_{\rm as}$ because, at terrestrial energy scales, meson DAs are much dilated with respect to $\varphi_{\rm as}$ and, concerning the kaon, somewhat skewed -- see, \emph{e.g}.,
Ref.\,\cite[Sec.\,3]{Roberts:2021nhw}.
Such dilation is consistent with analyses of existing pion+nucleon Drell-Yan data \cite{Xing:2023wuk}.
Furthermore, with the continuing development of novel algorithms, lQCD computations are also beginning to confirm these continuum predictions -- see, \emph{e.g}., Ref.\,\cite[Sec.\,8D]{Roberts:2021nhw}.
}

Employing the dilated DAs typical of modern analyses, Eq.\,\eqref{eq:pionFFUV} yields
{\allowdisplaybreaks
\begin{subequations}
\label{FPDAmod}
\begin{align}
\left. Q^2 F_\pi(Q^2)\right|_{Q^2 = 15\,{\rm GeV}^2}^{\varphi_{\rm dilated}} & = 0.16(1)\,{\rm GeV}^2, \\
\left. Q^2 F_K(Q^2)\right|_{Q^2 = 15\,{\rm GeV}^2}^{\varphi_{\rm dilated}} & = 0.21(1)\,{\rm GeV}^2\,,
\end{align}
\end{subequations}
\emph{viz}.\ enhancement by a factor $\approx 1.4$.  This explains some of the quantitative discrepancy between direct calculations and expectations based on Eq.\,\eqref{eq:pionFFUV}.  As noted elsewhere \cite{Chang:2013nia}, the remainder is likely explained by higher-order and higher-twist corrections to the hard scattering formula and, potentially, improvements to RL truncation.
}

Attempts to quantify the size of higher-order, higher-twist corrections to Eq.\,\eqref{eq:pionFFUV} do exist.  For instance, a recent analysis finds the corrections to be large and positive, but of uncertain magnitude \cite{Chai:2022srx}: estimates range from a multiplicative factor of $2-4$ on the values in Eq.\,\eqref{FPasy}.  Such magnitudes are far in excess of the corrections identified in Eq.\,\eqref{FPDAmod} and may require reconsideration because, in being so large, they question convergence of the expansions being used.

Regarding $\mathpzc Q2$, sound conclusions are nevertheless possible.
Namely, continuum and lattice analyses agree that at the scales for which terrestrial experiments are possible, $\varphi_{\rm as}$ is a poor approximation to the true results for $\pi$ and $K$ DAs.  Instead, $\varphi_{\pi,K}$ are both broad, concave functions -- see, \emph{e.g}., Ref.\,\cite[Fig.\,3.11]{Roberts:2021nhw}.
Thus, any attempt to estimate corrections to Eq.\,\eqref{eq:pionFFUV} and draw links with direct calculations of meson elastic electromagnetic form factors should begin by using these realistic (dilated and, for $K$, skewed) forms for meson leading-twist DAs.

\section{Summary and Perspective}
This study completes a unified, parameter-free treatment of the elastic electromagnetic form factors of the proton and $\pi$, $K$ mesons, delivering quantitative predictions for the meson form factors that cover the domain $0\leq Q^2/{\rm GeV}^2\leq 30$.
With the operation, construction, and planning of high-energy, high-luminosity facilities, one may expect these results to be tested in foreseeable experiments.

The results are also embedded within the context of a longstanding QCD prediction; namely, that positive-charge pseudoscalar meson elastic electromagnetic form factors, $F_{P^+}(Q^2)$, must fall faster than a dipole on some domain $Q^2 > Q_0^2\gg m_p^2$.  The multiplicative suppression is logarithmic: $[1/\ln Q^2]^{\gamma_F}$, $\gamma_F \approx 1.1$ at achievable high-energy scales.  Validating this prediction of scaling violation in meson form factors has been a longstanding experimental goal in hadron and particle physics, with even an expectation that the value of $\gamma_F$ may be accessible.

Herein, we have shown and stressed that since the form factors of such light mesons must be
monopole-like on $Q^2 \simeq 0$ and monotonically decreasing functions thereafter, then the existence of scaling violations means that $Q^2 F_{P^+}(Q^2)$ should exhibit a single maximum on $Q^2>0$.  The existence of a maximum is both a necessary and sufficient condition for the presence of scaling violations.

Our analysis indicates that, for charged $\pi$, $K$ mesons, the maximum of $Q^2 F_{P^+}(Q^2)$ lies in the neighbourhood $Q^2 \simeq 5\,$GeV$^2$.  Hence, proposed experiments are capable of locating the peak and, potentially, providing quantitative information on the value of $\gamma_F$.  These will be crucial steps toward ultimate verification of the QCD prediction that connects the hard scattering of leptons from mesons to expressions of emergent hadron mass in the Standard Model, via the effective charge and meson leptonic decay constant.

%
\medskip
\noindent\textbf{Acknowledgments}.
We are grateful to Z.-F.~Cui for valuable discussions.
Work supported by:
National Natural Science Foundation of China (grant no.\,12135007);
and STRONG-2020 ``The strong interaction at the frontier of knowledge: fundamental research and applications” which received funding from the European Union's Horizon 2020 research and innovation programme (grant agreement no.\ 824093).

\medskip
\noindent\textbf{Data Availability Statement}. This manuscript has no associated data or the data will not be deposited. [Authors' comment: All information necessary to reproduce the results described herein is contained in the material presented above.]

\medskip
\noindent\textbf{Declaration of Competing Interest}.
The authors declare that they have no known competing financial interests or personal relationships that could have appeared to influence the work reported in this paper.


\end{document}